\documentclass[aps, pra,twocolumn]{revtex4-1}
\usepackage{graphicx}
\usepackage[ansinew]{inputenc}
\usepackage{array}
\usepackage{color}
\usepackage{amsmath}
\usepackage{amsxtra}
\usepackage{amstext}
\usepackage{amssymb}
\usepackage{latexsym}
\newcommand{\bea}{\begin{eqnarray}}
\newcommand{\eea}{\end{eqnarray}}
\newcommand{\be}{\begin{equation}}
\newcommand{\ee}{\end{equation}}
\begin{document}

\title{Optomechanical back-action evading measurement without parametric instability}
\author{Steven K. Steinke,$^{1,2}$ K. C. Schwab,$^2$ and Pierre Meystre$^1$}
\affiliation{$^1$B2 Institute,  Department of Physics and College of Optical Sciences, University of Arizona, Tucson, AZ 85721, USA\\
$^2$Department of Applied Physics, Caltech, Pasadena, CA 91125, USA}

\date{\today}

\begin{abstract}
We review a scheme for performing a back-action evading measurement of one mechanical quadrature in an optomechanical setup. The experimental application of this scheme has been limited by parametric instabilities caused in general by a slight dependence of the mechanical frequency on the electromagnetic energy in the cavity. We find that a simple modification to the optical drive can effectively eliminate the parametric instability even at high intracavity power, allowing realistic devices to achieve sub-zero-point uncertainties in the measured quadrature.
\end{abstract}


\maketitle

\section{Introduction}

In combination with back-action evading techniques described below, squeezed states offer the potential for important applications in optomechanical precision force sensing~\cite{Caves1}, in particular achieving sensitivity with resolution below the zero-point fluctuations $x_\mathrm{zp}$ of the mechanical component of the sensor. Unfortunately, due to the typical thermal environment of a mechanical system, it is difficult to produce direct squeezing below the zero-point level in a mechanical oscillator. It is similarly challenging to ascertain its position to that level of accuracy in a single strong measurement. One possible solution is to continuously observe the position in, e.g., an optomechanical setup~\cite{reviews}, using light transmitted through a Fabry-P\'erot cavity to probe the motion of an oscillating end-mirror or using an equivalent microwave circuit with mechanically modulated frequency. Such a measurement process gradually reduces $\Delta x$. However, this leads to a problem as the oscillator rotates through phase space. Because measuring $\hat x$ increases the uncertainty of $\hat p$ and every quarter cycle $\hat x$ and $\hat p$ are exchanged, the maximum squeezing is limited to roughly that which could be produced in only one fourth of the oscillator period, even if technical noise is neglected. Hence, the potential squeezing is limited by measurement back-action.

A back-action evading measurement of the position of a membrane in a cavity optomechanical system was proposed as early as 1980 in Refs.~\cite{Braginsky1, Braginsky2}, who suggested driving the resonator with an input field resonant with the cavity frequency $\omega_c$, but modulated at the mirror frequency $\Omega_m$. In Fourier space, the field has oscillatory components at $\omega_c \pm \Omega_m$; hence, this scheme is often known as two-tone back-action evasion. By modulating the light field frequency at $\Omega_m$ the measurement effectively turns on and off as the system oscillates. This protocol thereby measures neither position nor momentum individually, but rather, one of the mechanical {\em quadratures}. Thus, while the measurement back-action still exists, it feeds only into the unmeasured quadrature and evades the measured one, leaving in principle no lower limit on the uncertainty one quadrature might reach.

A detailed quantum mechanical analysis is presented in Ref.~\cite{Clerk1}, and a generalized version exploiting intereference between additional tones is proposed in Ref.~\cite{Ono1}. The technique was originally explored for use in gravity wave detectors\cite{Braginsky3}, including an early approximation to stroboscopic position measurement\cite{Ono2}, but these experiments using massive oscillators were not designed to reach sensitivities near their zero-point levels.  More recently, experiments have been carried out in the quantum regime. These have reached sensitivities of $4 x_\mathrm{zp}$~\cite{Hertzberg}, $2.5 x_\mathrm{zp}$~\cite{Suh13}, and $1.4x_\mathrm{zp}$~\cite{SuhTLS}, but, thus far, no experiment has achieved sub-zero-point position sensitivity.

While two-tone back-action evasion is an elegant solution, experimental reality intervenes to place a rather restrictive limit on such a scheme. Because the envelope of the driving field oscillates at the mechanical frequency, the intracavity power oscillates at twice the mechanical frequency. Through indirect effects, this leads to the frequency of the mechanical oscillator becoming slightly modulated, with the modulation oscillating at twice the natural frequency. Such a frequency modulation produces a parametric instability, which will drive the system and greatly reduce the amount of squeezing possible. To accomplish the back-action evading measurement, high optical power and low mechanical dissipation are both critical, yet these factors both worsen the parametric instability.

The variety of ways in which this instability can arise is staggering. In the experiments mentioned above, it originated respectively from non-linear terms in the optomechanical coupling~\cite{Hertzberg}, cavity heating causing a thermal shift in the frequency of the mechanical element~\cite{Suh13}, and two-level systems in surface oxides acting as non-linear dielectrics~\cite{SuhTLS}. Thus, the common limitation of these experiments is a connection between mechanical frequency and the cavity energy, $\delta\Omega_m\propto E$. In the two-tone scheme, this connection seems to lead inevitably to parametric instability.

However, it is in principle possible to work around this particular limitation. There are two critical features needed to avoid the parametric instability while performing the back-action evading measurement:
\begin{enumerate}
\item The power in the cavity must not oscillate at twice the mechanical frequency.
\item The probe light must couple only to a single quadrature of motion.
\end{enumerate}
The first of these requirements prevents the instability, as we will discuss in more detail in the next section. The latter requirement prevents measurement back-action from affecting the measured quadrature. Small deviations from this requirement will reduce the ultimate sensitivity of the measurement, but do not preclude sensitivities below the zero-point level, as shown in Ref.~\cite{Clerk1}.

We will verify in this paper that these two criteria may be pursued somewhat independently of each other. In the next section, we show that the ideal envelope for the intracavity field is a square wave, which satisfies the first criterion above. While experimental constraints prevent the realization of a perfect square wave, we show that it is sufficient to add a single additional drive tone. The rest of the paper is devoted to proving that the second criterion will also be satisfied. We review in Section III the optomechanical system used for the measurement, including a general optical driving term and outlining the main steps in extending the results of Ref.~\cite{Milburn1} for such a drive. Using reasonable simplifications, in Section IV we then derive a general master equation for the conditional evolution of the mechanics under continuous observation and show in Section V that, for the desired purpose, it can be reduced to the master equation of Ref.~\cite{Clerk1}, thereby completing the demonstration of the consistency of the two criteria. Finally, Section VI is a summary and outlook.

\section{Field Envelope}
We model a generic cavity optomechanical system as a single-mode optical or electrical resonator with resonant mode frequency $\omega_c$ whose radiation pressure drives an harmonically confined end-mirror or capacitor plate with natural oscillation frequency $\Omega_m$. The cavity is driven on-resonance but modulated with a field envelope $\alpha(t)$. In the microwave domain, this cavity field is often achieved through the application of multiple tones in order to avoid the low frequency phase noise of available sources.

We proceed by first asking what specific drive scheme can be used to avoid the mechanical parametric instability. To recap, this instability is due to a generic coupling arising through a variety of mechanisms between the mechanical frequency and the electromagnetic energy in the cavity which oscillates proportional to $|\alpha(t)|^2$. If only two tones are used to drive the system at $\omega_c\pm\Omega_m$, the cavity power then oscillates at $2\Omega_m$. Via device-dependent mechanisms, this induces a shift in the mechanical frequency,
\be
\label{freqshiftproblem}
\Omega^\prime_m(t) = \Omega_m + \delta\Omega_m\cos2\Omega_mt.
\ee
In a well-made device, the shift $\delta\Omega_m$ will be much less than $\Omega_m$, though the quality factor $Q$ will also be high. Hence the mechanical damping will be weak. The onset of parametric instability begins when the criterion for the fractional frequency shift
\be
\label{minfreqshift}
\frac{\delta\Omega_m}{\Omega_m} > \frac{1}{Q}
\ee
is satisifed\cite{Xie1}. Similar parametric resonances occur if the mechanical frequency oscillates at subharmonics of $2\Omega_m$, but our scheme avoids these. Furthermore, the required fractional frequency shift is of order $Q^{-1/n}$ for the $n$th subharmonic, rendering all but the first instability irrelevant for our purposes. However, even in carefully engineered devices, the limit in Eq.~\eqref{minfreqshift} is reached at low pump power, reducing the maximum possible accuracy of the back-action evading measurement to above the zero point level\cite{SuhTLS}.

We will show in the following sections that the measurement of the membrane motion can be performed in such a way that it is dominated by the two tones at $\omega_c\pm\Omega_m$. Therefore, we are more or less free to add additional fields to the cavity, as long as they are not near those frequencies, without affecting the measurement-based squeezing. We can in that way cancel out the oscillations of the energy in the cavity at $2\Omega_m$. Of course, this will add oscillations at higher harmonics, but parametric resonance only occurs at {\em  subharmonics} of $2\Omega_m$\cite{Xie1}.

\begin{figure}
\begin{center}
\includegraphics[width=0.5\textwidth]{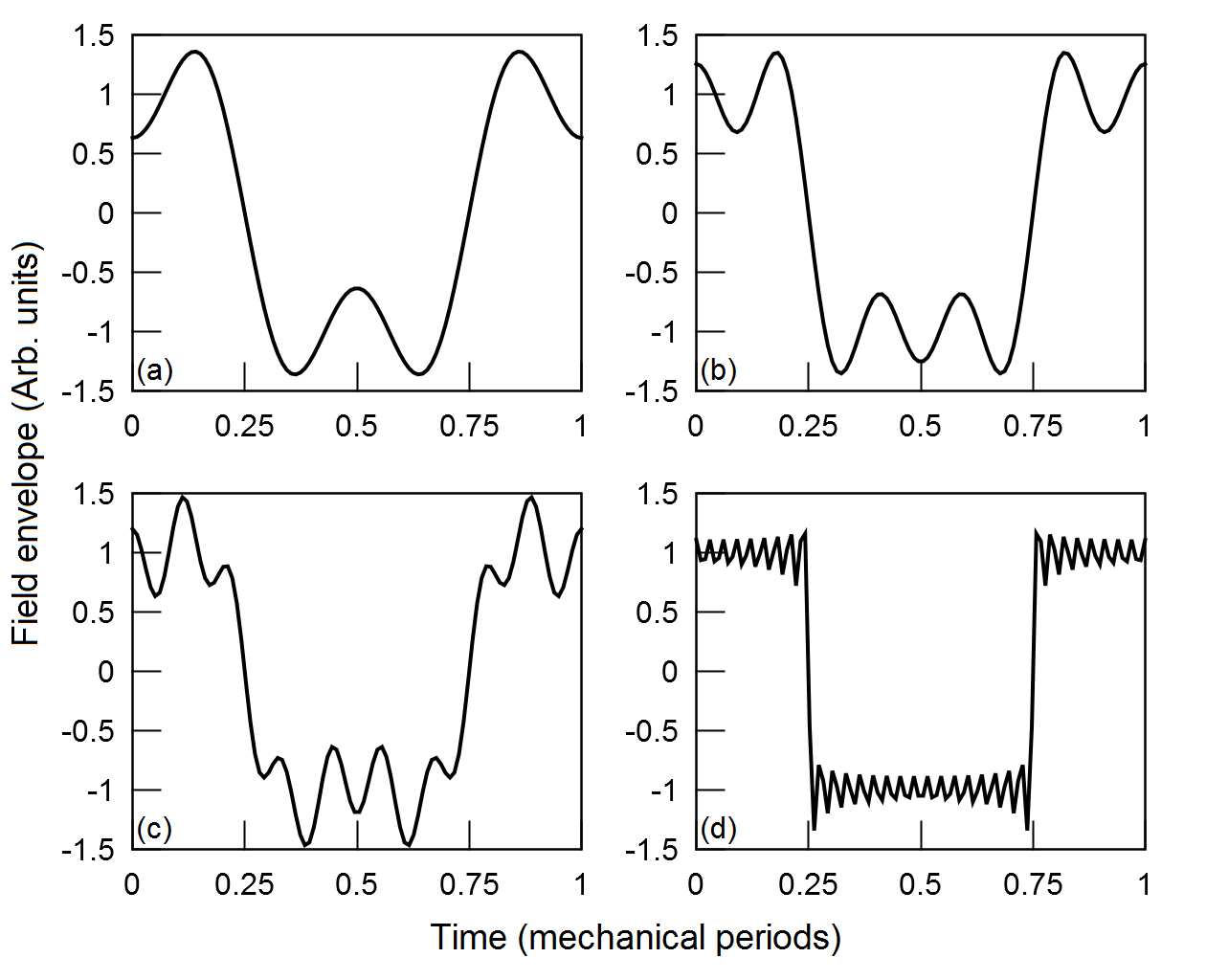}
\end{center}
\caption{\label{multiplot}The field envelope $\alpha(t)$ needed to cancel higher harmonic oscillations in the intracavity energy $|\alpha(t)|^2$, using (a) 1, (b) 2, (c) 4, and (d) 16 additional drive tones.}
\end{figure}

Consider specifically a field envelope, i.e., the electromagnetic field in a frame rotating at $\omega_c$, with a single added drive tone at $3\Omega_m$, 
\be
\label{omega3}
\alpha(t)\propto\cos\Omega_m t + \mu e^{i(3\Omega_mt+\Phi)}.
\ee
The energy $E$ in the cavity is proportional to $|\alpha|^2$,
\be
\label{omega246}
E(t) \propto \frac{1}{2}+\mu^2+A_2\cos \left(2\Omega_mt+\Phi_2\right) +\mu\cos\left(4\Omega_mt+\Phi\right),
\ee
with
\bea
\label{a2phi2}
A_2 &=& \sqrt{\frac{1}{4}+\mu\cos\Phi+\mu^2},\nonumber\\
\Phi_2 &=&\arctan\frac{2\mu\sin\Phi}{1+2\mu\cos\Phi}.
\eea
Thus, by adding the additional $3\Omega_m$-detuned tone with $\mu=1/2$ and $\Phi=\pi$, the energy oscillating at $2\Omega_m$ is completely redirected via interference into the DC and $4\Omega_m$ terms. For small deviations from ideality, i.e., $\mu=1/2+\delta\mu$, $\Phi=\pi+\delta\Phi$, the relative amplitude of the energy oscillation at $2\Omega_m$ is
\be
\label{nonidealcase}
A_2=\sqrt{(\delta\mu)^2+\frac{(\delta\Phi)^2}{4}},
\ee
which still represents a significant reduction in the $2\Omega_m$ Fourier component. In the next section, we will discuss the drive needed to produce such the cavity field of Eq.~\eqref{omega3}. If desired, one could continue to eliminate the higher harmonics of the energy oscillation by adding additional phase-locked sources detuned from the cavity by $\pm(2n+1)\Omega_m$, for increasing integer $n$. If we extend the series in Eq.~\eqref{omega3}, use tones of equal strength at each pair of red- and blue-detuned sidebands (i.e. cosines rather than complex exponentials) so that $\alpha(t)$ is real, and repeat the cancellation procedure of Eq.~\eqref{omega246} for these higher harmonics, we find that $\alpha(t)$ converges to a square wave envelope. Solving for such field envelopes requires finding roots of higher order polynomials for higher harmonics, so it must be done numerically. We plot several such solutions in Fig.~\ref{multiplot}. It is also intuitively easy to verify that the square wave has the desired properties: it is a function that oscillates at the same fundamental frequency as the mechanics, thereby permitting the measurement of a single quadrature, but, when squared, it is simply a constant. Thus, the energy in the cavity remains constant, and there is no resulting oscillatory mechanical frequency shift at any harmonic of the mechanical motion. In terms of the field itself rather than the envelope, this corresponds to a $\pi$ phase shift every half mechanical period.
However, adding numerous phase-locked tones may be technically difficult, and so we examine the effectiveness of a single added $3\Omega_m$-detuned tone. The energy in the cavity, and hence the mechanical frequency, now has a Fourier component, aside from the DC term, at $4\Omega_m$. It is, hypothetically speaking, still possible to excite a parametric instability in this case; however, the required fractional frequency shift in the mechanics to excite a parametric instability is of order 1, i.e., the mechanical frequency would have to vary wildly during the course of the experiment~\cite{Xie1}. Such large frequency shifts are simply not seen in these devices; instead, the fractional shifts are of order less than a percent~\cite{Hertzberg,Suh13,SuhTLS}. To contrast, the fractional frequency shift required to excite a parametric instability is only $\frac{1}{Q}$ when the oscillations in frequency occur at $2\Omega_m$. A shift of this magnitude is seen experimentally, such as in Ref.~\cite{SuhTLS}, where the observed frequency shift of $\sim$ 50 Hz agrees quite well with the predicted onset of parametric instability. Based on these observations, we therefore find that a single additional tone is sufficient to eliminate the instability.

\section{Model}
Having now satisfied the first key criterion listed in the introduction we turn to the demonstration that, under realistic conditions, the probe light couples predominantly to a single quadrature of motion. Our starting point is the generic optomechanical Hamiltonian, see e.g. Ref.~\cite{Clerk1}
\bea
\label{hamiltonian} 
H &=& \hbar(\omega_c - G \hat x)\left [\hat a^\dagger \hat a - \langle \hat a^\dagger \hat a\rangle(t)\right ]+\hbar \Omega_m \hat c^\dagger \hat c \nonumber \\
&+&i\hbar\sqrt{\kappa}(b^*_\mathrm{in}(t)\hat a-b_\mathrm{in}(t)\hat a^\dag),
\eea
that describes the dynamics of a single-mode Fabry-P{\'e}rot resonator with an oscillating, harmonically bound  end-mirror driven by radiation pressure. Here $\hat a$ is the annihilation operator of the cavity mode and $\kappa$ its decay rate, $b_\mathrm{in}$ is the amplitude of the (classical) driving field, and $\hat x = x_\mathrm{zp}(\hat c+\hat c^\dag)$ is the displacement of the mirror resulting from radiation pressure. $G$ is the single photon optomechanical coupling. Finally $x_\mathrm{zp}^2$ is the position variance of the membrane in its ground state. 

We assume that the drive is resonant with the cavity but periodically modulated at the mechanical frequency,
\bea
\label{bin}b_\mathrm{in}(t) &=& e^{-i\omega_c t}\sum_{\ell=-\infty}^{\infty}\beta_\ell e^{i\Omega_m \ell t}.
\eea
This differs from the scheme of Refs.~\cite{Braginsky1,Clerk1} in which only $\beta_{\pm 1} \ne 0$. We further assume that the mechanical element is thermally damped at a rate $\gamma$
due to its contact with a reservoir of temperature $T$. Thus its equilibrium thermal phonon occupancy is given
by the Bose-Einstein distribution,
\begin{equation}
\bar n = \frac{1}{\exp(\hbar\Omega_m / k_B T)-1}.
\end{equation}
However, we assume that $k_B T \ll \hbar\omega_c$ so that we can neglect thermal photons. 

The effects of thermal coupling and the cavity decay are encapsulated by the master equation for the composite
density matrix $\rho$ of the oscillator-field system,
\be
\label{master}
\frac{d \rho}{dt} = \frac{i}{\hbar}[\rho,H] + \frac{\kappa}{2}\mathcal{D}[\hat a]\rho
+\frac{\gamma}{2}\left ((\bar n+1)\mathcal{D}[\hat c]\rho+\bar n\mathcal{D}[\hat c^\dag]\rho\right ),
\ee
where the dissipation super-operator $\mathcal{D}$ is defined as
\begin{equation}
\mathcal{D}[\hat o]\rho = 2\hat o\rho \hat o^\dag - \hat o^\dag \hat o\rho - \rho \hat o^\dag \hat o.
\end{equation}
We also work in the good cavity limit, specifically assuming a separation of scales
\begin{equation}
\label{separationofscales}
\gamma \ll \kappa \ll \Omega_m \ll \omega_c.
\end{equation}
The only of the above inequalities that is not trivially satisfied in the typical experimental environment is $\kappa \ll \Omega_m$. In the optical domain, even cavities with very high finesse have $\kappa$ of order $\sim$ 10 MHz due to the high frequency of light. Working in the microwave regime eases this requirement somewhat; typical parameters in a microwave device are $\omega_c = 2\pi\times5.3$~GHz, $\Omega_m=2\pi\times3.7$~MHz, $\kappa=2\pi\times260$~kHz, and $\gamma=2\pi\times50$~Hz~\cite{SuhTLS}.

We now simplify the master equation by moving into the rotating frame for both the optical mode and the mechanical oscillator, and then displacing $\hat a$ by a classical mean value $\alpha(t)$ -- ultimately, the field envelope of the previous section -- where the time dependence here is due to the modulated nature of the drive. Mathematically, this is equivalent to applying several unitary transformations to the density matrix in succession
\be
\rho^\prime = WVU\rho U^\dag V^\dag W^\dag,
\ee
where
\begin{eqnarray}
U&=&\exp(i\Omega_m \hat c^\dag \hat c t),\nonumber \\
V&=&\exp(i\omega_c \hat a^\dag \hat a t),\nonumber \\
W&=&\exp[\alpha(t)\hat a^\dag - \alpha^*(t)\hat a].
\end{eqnarray}
Though $W(t)$ does not typically commute with $W(t^\prime)$, this fact ultimately contributes only an irrelevant net global phase to the evolution. Identifying $\langle \hat a^\dag \hat a\rangle = |\alpha|^2$ and noting that in the rotating frame the mean intracavity field satisfies 
\begin{equation}
d\alpha/dt+\kappa\alpha/2=\sqrt{\kappa}b_\mathrm{in}e^{i\omega_ct}.
\end{equation}
we find readily
\bea
\label{meanalpha}\alpha(t)&=&\sum_{\ell=-\infty}^{\infty}\alpha_\ell e^{i\Omega_m\ell t},\nonumber \\
\alpha_\ell \ &=& \frac{\sqrt{\kappa}\beta_\ell}{i\Omega_m\ell+\kappa/2}.
\eea
We can combine these formulae with the results of the previous section to provide the exact form of the needed drive for the $3\Omega_m$-detuned tone. Specifically, if the first red and blue $\Omega_m$-detuned sidebands are pumped with amplitude and phase given by
\be
\beta_{\pm 1} = \left (\frac{\kappa}{2}\pm i\Omega_m\right)B,
\ee
then the third sideband should be pumped with amplitude and phase
\be
\beta_3 = -\left (\frac{\kappa}{2}+3i\Omega_m\right)B.
\ee

After the above unitary transformations and substitutions, the Hamiltonian governing the evolution of the system in the primed frame is
\be
\label{hprime}
H^\prime = g\left(\hat ce^{-i\Omega_mt}+\hat c^\dag e^{i\Omega_mt}\right)
\left(\alpha(t) \hat a^\dag+\alpha^*(t) \hat a+\hat a^\dag \hat a\right),
\ee
where  $g = Gx_\mathrm{zp}$. The dissipative terms remain unchanged, except for the replacement of $\rho$ with
$\rho^\prime$.
In subsequent calculations, we neglect the quadratic $\hat a^\dag \hat a$ term in the Hamiltonian (\ref{hprime}), as it is of order unity in size. On the other hand, the linear terms are multiplied by the classical mean field, $\alpha(t)$, which is the square root of the mean intracavity photon number, typically $10^6$ or more for back-action evading experiments, so these will dominate the evolution dynamics.

\section{Optomechanical Master Equation}
We now turn to the issue of measurement of the system. Our derivation follows the approach of Ref.~\cite{Milburn1}, generalized to a local oscillator with a time-dependent amplitude or phase. 

By using a homodyne detection scheme, it is possible to make a measurement of one quadrature of the cavity field. The output field from the cavity is 
\begin{equation}
\hat b_\mathrm{\rm out} = b_\mathrm{\rm in} + \sqrt{\kappa}\hat a
\end{equation}
and the field reaching the detector is $B_\mathrm{\rm lo} + \hat b_{\rm out}$, where $B_{\rm lo}$ is the additional local oscillator. We can fold all the classical contributions together into a ``net'' local oscillator strength given in the rotating frame by
\begin{equation}
B_\mathrm{net} = (B_\mathrm{lo}+b_\mathrm{in})e^{i\omega_ct}/\sqrt{\kappa}-\alpha(t) \equiv B(t)e^{i\phi(t)}
\end{equation}
where  $\phi$ is relative phase between the local oscillator and the ouput field. Note that if $B_\mathrm{lo}$ is sufficiently large, the additional terms are negligible. For a detector of efficiency $\eta$ the detected photocurrent $I$ (i.e., the measurement record) is given by
\be\label{meas}
Idt = \left(B^2+B\langle \hat a e^{-i\phi} + \hat a^\dag e^{i\phi}\rangle\right)\eta\kappa dt + B\sqrt{\eta\kappa}dW,
\ee
where $W(t)$ is a Wiener process. That is, $\xi(t) = dW/dt$ is Gaussian white noise and $(dW)^2 = dt$. The dynamical effects of this measurement on the density matrix can be computed and yield the conditional stochastic master equation (SME)~\cite{Milburn1}
\bea
\label{SME}
\frac{d \rho_c}{dt} &=& \frac{i}{\hbar}[\rho_c,H^\prime] + \frac{\kappa}{2}\mathcal{D}[\hat a]\rho_c \nonumber \\
&+&\frac{\gamma}{2}\left [(\bar n+1)\mathcal{D}[\hat c]\rho_c+\bar n\mathcal{D}[\hat c^\dag]\rho_c\right ] \\
&+& \left (\hat a\rho e^{-i\phi}+\rho \hat a^\dag e^{i\phi}-\langle \hat a e^{-i\phi}+\hat a^\dag e^{i\phi}\rangle\rho\right )\sqrt{\eta\kappa}\xi(t).\nonumber
\eea

\section{Reduced Master Equation And Measurement}
In order to proceed, we now adiabatically eliminate the light field. This can be achieved by working in the weak-coupling limit. Specifically, we assume that the optomechanical interaction strength, approximately given by $g|\alpha(t) \langle\hat c \rangle |$, is much less than the cavity decay rate $\kappa$. This has multifold advantages: First, we can accurately make the rotating wave approximation (RWA), thereby removing the explicit time dependence of the interaction Hamiltonian and simplifying it to the form given below in Eqs.~\eqref{hrwa1} and~\eqref{hrwa2}.
Second, we find rate equations for those components of the density matrix needed to trace out the light field and derive an effective master equation for the mechanical subsystem alone. Those terms not involved in the trace, i.e. the off-diagonal terms, are taken to adiabatically follow the other terms due to the dominance of the decay rate $\kappa$. Because similar calculations have been reported thoroughly elsewhere, including in Ref.~\cite{Milburn1}, we do not reproduce all intermediate details here. 

Our first step is to make the RWA, after which we are left with the interaction
\be
\label{hrwa1}H^\prime = g\left(\hat C_1^\dag \hat a+\hat C_1\hat a^\dag\right),
\ee
where 
\be
\hat C_1 = \alpha_1\hat c+\alpha_{-1}\hat c^\dag
\ee
 and
 \be
 [\hat C_1,\hat C_1^\dag] = |\alpha_1|^2-|\alpha_{-1}|^2.
 \ee
If the red and blue sidebands of the field detuned at $\pm\Omega_m$ are balanced in intensity, $\alpha_1 = Ae^{i\theta} = \alpha_{-1}^*$, with $A$ real, and $\hat C_1$ simplifies to a constant times the Hermitian mechanical quadrature operator
\be
\hat Q = \frac{1}{\sqrt{2}}\left(e^{i\theta}\hat c+e^{-i\theta}\hat c^\dag\right).
\ee
Namely,
\be
\hat C_1 = A\sqrt{2}\hat Q,
\ee
and the Hamiltonian ultimately reduces to,
\be
\label{hrwa2}
H^\prime = gA\sqrt{2}\left(\hat a + \hat a^\dag\right)\hat{Q}.
\ee

The next step is to define the small parameter
\be
\epsilon = \frac{gA}{\kappa}.
\ee
Because of the dominant effect of dissipation on the dynamics of the intracavity light field it always remain near its equilibrium, absent the optomechanical interaction. In the displaced frame, this is the ground state. Therefore, we can expand the density matrix in the (displaced) optical Fock basis as
\bea
\label{expansion}
\rho &=& \rho_{00}|0\rangle\langle 0| + \epsilon(\rho_{01}|0\rangle\langle 1|+\mathrm{H.c.})\nonumber \\
&+&\epsilon^2 \rho_{11}|1\rangle\langle 1|+\epsilon^2(\rho_{02}|0\rangle\langle 2|+\mathrm{H.c.})
+\mathcal{O}(\epsilon^3).
\eea
The adiabatic elimination proceeds by substituting equation \eqref{expansion} into the stochastic master equation~\eqref{SME}. We can then obtain a dimensionless version of the SME by rescaling time and the Wiener increment, $d\tau = \kappa dt, dw = \sqrt{\kappa}dW$. (Note that $dw/d\tau$ is still Gaussian white noise because $(dw)^2 = d\tau$.) Adiabatically eliminating the off-diagonal terms and truncating after order $\epsilon^2$ yields the relations
\be
\label{offdiag}
\rho_{02} = 2i\rho_{01}\hat Q,\hspace{24pt}\rho_{01} = 2i\sqrt{2}\rho_{00}\hat Q,
\ee
and the equations of motion for the diagonal terms are (again, to order $\epsilon^2$)
\bea
\label{rho00}
&&d\rho_{00}= \frac{\gamma}{2\kappa}\left [(\bar n+1)\mathcal{D}[\hat c]\rho_{00}+\bar n\mathcal{D}[\hat c^\dag]\rho_{00}\right ]d\tau \nonumber\\
&&+\epsilon^2\left[i\sqrt{2}\left(\rho_{01} \hat Q-\hat Q\rho^\dag_{01}\right)+\rho_{11}\right ] d\tau\nonumber \\
&&+\left[\rho_{01}e^{i\phi}+\rho^\dag_{01}e^{-i\phi}-\rho_{00}\mathrm{tr} \left(\rho_{01}e^{i\phi}+\rho^\dag_{01}e^{-i\phi}\right)\right ]\epsilon dw,\nonumber \\
\label{rho11}
&&d\rho_{11}=\frac{\gamma}{2\kappa}\left [(\bar n+1)\mathcal{D}[\hat c]\rho_{11}+\bar n\mathcal{D}[\hat c^\dag]\rho_{11}\right ] d\tau \nonumber\\
&&-\left [ (i\sqrt{2}\left(\hat Q\rho_{01}-\rho^\dag_{01}\hat Q\right)+\rho_{11}\right ]d\tau,
\eea
where the trace in Eq.~\eqref{rho00} is over the mechanical degree of freedom. Though $\rho_{11}$ is a small term (of order $\epsilon^2$), its inclusion simplifies the computation of the final master equation for the mechanics.

Tracing over the optical degree of freedom is equivalent to finding an expression for $d(\rho_{00}+\epsilon^2\rho_{11})$, which is done easily by substitution of Eq.~\eqref{offdiag} into Eqs.~\eqref{rho00}. After substituting back in $t$ and $W$, this yields the stochastic master equation for the conditional density matrix of the mechanical subsystem,
\bea
\label{finalME}
\frac{d \rho_m}{dt} &=& -k[\hat Q,[\hat Q,\rho_m]] \nonumber \\
&+& i\sqrt{2\eta k}(e^{i\phi}\rho_m \hat Q-e^{-i\phi}\hat Q\rho_m - 2i\sin\phi\langle \hat Q\rangle\rho_m)\xi(t)\nonumber \\
&+&\frac{\gamma}{2}((\bar n+1)\mathcal{D}[\hat c]\rho_m+\bar n\mathcal{D}[\hat c^\dag]\rho_m),
\eea
where $k = 4g^2 A^2/\kappa$.

The relative phase of $\alpha_1$ and $\alpha_{-1}$ defines which quadrature of motion is measured, and without loss of generality we can take $\theta = 0$ so that $\hat Q = \hat X$. We represent the orthogonal quadrature by $\hat Y$. On the other hand, Eq.~\eqref{finalME} makes explicit the importance of the relative phase $\phi$ between the local oscillator used for detection and the classical driving field. If they are $\pi/2$ out of phase, then maximum measurement strength is achieved. By contrast, if they are in phase ($\phi=0$), then the measurement acts as a stochastic unitary drive of the system, i.e., a random force displacing the mechanics. In this case, the light quadrature being measured is $\hat a+\hat a^\dag$, which amounts to replacing that term in the Hamiltonian with a fluctuating, semi-classical value.

We now take $\phi=\pi/2$. This step is included for completeness, as it allows us to reproduce the master equation and then the key results on conditional squeezing of Ref.~\cite{Clerk1}, 
\bea
\label{finalfinalME}
\frac{d \rho_m}{dt} &=& -k[\hat X,[\hat X,\rho_m]]  - \sqrt{2\eta k}(\rho_m \hat X+\hat X\rho_m -2\langle \hat X \rangle)\xi(t)\nonumber \\
&+&
\frac{\gamma}{2}\left ((\bar n+1)\mathcal{D}[c]\rho+\bar n\mathcal{D}[c^\dag]\rho\right ).
\eea

The equations for the {\em conditioned} mean values, variances, and covariance ($C=\langle \hat X\hat Y+\hat Y\hat X\rangle/2-\langle \hat  X\rangle\langle \hat Y\rangle$) are readily derived under a Gaussian state ensatz:
\bea
\frac{d}{dt}\langle \hat X\rangle&=&-\frac{\gamma}{2}\langle X\rangle-\sqrt{K}V_X\xi,\\
\frac{d}{dt}\langle \hat Y\rangle&=&-\frac{\gamma}{2}\langle Y\rangle-\sqrt{K}C\xi,\\
\frac{dV_X}{dt} &=& -K V_X^2 -\gamma\left(V_X-\bar n-\frac{1}{2}\right),\\
\frac{dV_Y}{dt} &=& -K C^2 + 2k - \gamma\left(V_Y-\bar n-\frac{1}{2}\right),\\
\frac{dC}{dt} &=& -K V_X C -\gamma C,
\eea
where we have introduced the scaled measurement strength $K = 8\eta k$. These variances approach steady state values. Of particular interest is the variance of $\hat X$
\bea
V_X&=&\sqrt{\frac{\gamma}{2K}\left(2 \bar n + 1+\frac{\gamma}{2K}\right)}
-\frac{\gamma}{2K},
\eea
which approaches 0 for sufficiently large $K$ or small $\gamma$; the uncertainty relations are maintained by a concommitant increase in $V_Y$. As mentioned above, these results agree exactly with those of Ref.~\cite{Clerk1}, despite the modification to the optical drive. The physical reason these distinct schemes produce the same results is fairly straightforward. The added light harmonics couple weakly and oscillate three times faster than the mechanical frequency, so the force they exert on the mechanical element will average out to zero over each mechanical period. Because our modified system produces all the same results from this point, we will omit the extensive additional calculations on the use of feedback to promote conditional squeezing into unconditioned, or ``real'', squeezing.

Summarizing, then, we have shown that, provided that the RWA can be invoked (i.e., the good cavity limit of Eq.\eqref{separationofscales} is satisfied), only the contribution of the first sideband, $\ell=\pm 1$, contributes significantly to the homodyne detection signal, demonstrating that the probe field couples predominantly to a single quadrature of motion. With this, both criteria discussed in the introduction are satisfied, demonstrating the viability of the proposed scheme for eliminating the parametric instability.

\section{Conclusions}
We have thus extended the proposal to perform optomechanical back-action evading measurements to the case of a multi-tone drive scheme. Specifically, by simply adding a third tone detuned from the cavity by $3\Omega_m$ with appropriate amplitude and phase, we push the oscillations in cavity energy to higher harmonics of the mechanical frequency, which in turn do not contribute to the parametric instability. We then reproduced, in the appropriate good-cavity but weakly coupled regime ($gA\ll\kappa\ll\Omega_m$), the desired back-action evading measurement of a single mechanical quadrature. Because these instabilities appear in such diverse experimental settings, we hope this modification will prove useful in reaching sub-zero-point position sensitivities deep in the quantum regime. Indeed, preliminary experimental results from our collaboration show squeezing in a microwave optomechanical device below the zero-point level; we will confirm and refine these results in the coming months.

There is still room to extend the calculations presented above. For instance, it may be instructive to include the frequency modulating terms explicitly in the Hamiltonian, $H_\mathrm{parametric} \propto a^\dag ac^\dag c$. In addition, we can go beyond the rotating wave approximation to examine the back-action contributed by the rapidly oscillating terms in the Hamiltonian. While small, this contribution is non-zero, and could eventually impose a lower limit once other technical obstacles are overcome. Finally, measurements of the output field other than homodyne detection should be considered to increase the applicability of the scheme even further.

\begin{acknowledgments}
This work was supported by the DARPA QuASAR program through a grant from AFOSR and the DARPA ORCHID program through a grant from ARO, the US Army Research Office, and by the NSF. KCS acknowledges funding provided by the Institute for Quantum Information and Matter, an NSF Physics Frontiers Center with support of the Gordon and Betty Moore Foundation through Grant GBF1250.
\end{acknowledgments}


\begin{thebibliography}{10}
\bibitem{Caves1}
C. M. Caves, K. S. Thorne, R. W. P. Drever, V. D. Sandberg, M. Zimmermann, Rev. Mod. Phys. {\bf 52}, 341 (1980)
\bibitem{reviews}
For recent reviews of cavity and quantum optomechanics, see T. J. Kippenberg, K. J. Vahala, Science {\bf 321} 1172 (2008); M. Aspelmeyer {\it et al.}, J. Opt. Soc. Am. B {\bf 27}, A189 (2010); F. Marquardt, S. M. Girvin, Physics {\bf 2}, 40 (2009); M. Aspelmeyer, P. Meystre, K. C. Schwab, Physics Today {\bf 65}, 29 (2012); P. Meystre, Ann. der Physik {\bf 525}, 215 (2013); D. M. Stamper-Kurn, arXiv:1204.4351, to appear in {\em Cavity Optomechanics}, edited by M. Aspelmeyer, T. Kippenberg, F. Marquardt, Springer Verlag; M. Aspelmeyer, T. J. Kippenberg, F. Marquardt, arXiv:1303.0733, to be published in Rev. Mod. Phys.
\bibitem{Braginsky1}
V. B. Braginsky, Y. I. Vorontsov, K. S. Thorne, Science {\bf 209}, 4456 (1980).
\bibitem{Braginsky2}
V. B. Braginsky, F. Ya. Khalili, {\em Quantum Measurement,} Cambridge University Press, Cambridge, UK, (1992).
\bibitem{Braginsky3}
V. B. Braginsky, F. Ya. Khalili, Rev. Mod. Phys. {\bf 68}, 1 (1996).
\bibitem{Clerk1}
A. A. Clerk, F. Marquardt, K. Jacobs, New J. Phys. {\bf 10}, 095010 (2008).
\bibitem{Ono1}
R. Onofrio, F. Bordoni, Phys. Rev. A {\bf 43}, 2113 (1991).
\bibitem{Ono2}
L. E. Marchese, M. F. Bocko, R. Onofrio, Phys. Rev. D {\bf 45}, 1869 (1992).
\bibitem{Hertzberg}
J.B. Hertzberg, T. Roucheleau, T. Ndukum, M. Savva, A. A. Clerk, K. C. Schwab, Nature Physics {\bf 6}, 213 (2009).
\bibitem{Suh13}
J. Suh, M.D. Shaw, H.G. LeDuc, A.J. Weinstein, K.C. Schwab, Nano Lett {\bf 12}, 6260 (2012).
\bibitem{SuhTLS}
J. Suh, A.J. Weinstein, K.C. Schwab, Appl. Phys. Lett. {\bf 103}, 052604 (2013).
\bibitem{Milburn1}
H. M. Wiseman, G. J. Milburn, Phys. Rev. A {\bf 47}, 642 (1993).
\bibitem{Xie1}
W. Xie, {\em Dynamic Stability of Structures,} Cambridge University Press, New York, NY, USA (2006).
\end{thebibliography}
\end{document}